\documentclass[ reprint,
 amsmath,amssymb, aps,  prl]{revtex4-1}

\usepackage{amsmath,amsfonts,amstext,amssymb,mathrsfs}
\usepackage{graphicx, xcolor}
\usepackage[colorlinks=true,linkcolor=navy,urlcolor=navy,
citecolor=navy]{hyperref}
\RequirePackage{color}
\definecolor{navy}{RGB}{0,0,202} 
\usepackage{dcolumn}
\usepackage{bm}
\usepackage{multirow}
\usepackage{comment}
\usepackage{tikz}
\newcommand{\D}{\mathrm{d}}

\renewcommand{\O}{\mathcal{O}}
\newcommand{\<}{\langle}
\renewcommand{\>}{\rangle}
\newcommand{\bs}[1]{\boldsymbol{#1}}
\newcommand{\nn}{\nonumber}
\newcommand{\lla}{\langle \! \langle}
\newcommand{\rra}{\rangle \! \rangle}

\newcommand{\lwick}{:\!}
\newcommand{\rwick}{\!:}

\begin{document}

\title{\texorpdfstring{Conformal $\bs{n}$-point functions in momentum space}{Conformal n-point functions in momentum space}}

\author{Adam Bzowski}
\email{adam.bzowski@physics.uu.se}
\affiliation{Department of Physics and Astronomy, Uppsala University,
751 08 Uppsala, Sweden.}

\author{Paul McFadden}
\email{paul.l.mcfadden@newcastle.ac.uk}
\affiliation{School of Mathematics, Statistics \& Physics, Newcastle University, 
Newcastle NE1 7RU, U.K.} 

\author{Kostas Skenderis}
\email{k.skenderis@soton.ac.uk}
\affiliation{STAG Research Center \& Mathematical Sciences, 
University of Southampton,
Highfield, 
Southampton 
SO17 1BJ, U.K.}


\begin{abstract}

We present a Feynman integral representation for the general momentum-space scalar $n$-point function in any conformal field theory. This representation solves the conformal Ward identities
and features an arbitrary function of $n(n-3)/2$ variables which play the role of momentum-space conformal cross-ratios. 
It involves $(n-1)(n-2)/2$ integrations over momenta, 
with the momenta running over the edges of 
an $(n-1)$-simplex. We provide the details in the simplest non-trivial  case (4-point functions), and for this case
we identify  values of the operator and spacetime dimensions for which singularities arise leading to anomalies and beta functions, and discuss
several 
illustrative examples from perturbative quantum field theory and holography. 

\end{abstract}


\maketitle

\section{\hspace{-3mm}I. \hspace{1mm} Motivation}

The structure of correlation functions in a conformal field theory (CFT) 
is highly constrained by conformal symmetry. 
It has been known since the work of Polyakov \cite{Polyakov:1970xd, DiFrancesco:1997nk} that the most general 4-point function of scalar primary operators $\O_{\Delta_j}$,  each of dimension $\Delta_j$, takes the form
\begin{align}
& \< \O_{\Delta_1}(\bs{x}_1) \O_{\Delta_2}(\bs{x}_2) \O_{\Delta_3}(\bs{x}_3) \O_{\Delta_4}(\bs{x}_4) \> \nn\\[0.5ex]& \qquad 
= f(u, v) \prod_{1 \leq i < j \leq 4} x_{ij}^{2 \delta_{ij}}, \label{x4point}
\end{align}
where  $x_{ij} = | \bs{x}_i - \bs{x}_j |$ are the coordinate separations and
\begin{equation} \label{params}
2 \delta_{ij} = \frac{\Delta_t}{3} - \Delta_i - \Delta_j, \qquad \Delta_t = \sum_{i=1}^4 \Delta_i.
\end{equation}
The 4-point function is thus determined up to an arbitrary (theory-specific) function $f$ of the two conformal cross-ratios, 
\begin{equation} \label{std_uv}
u = \frac{x_{13}^2 x_{24}^2}{x_{14}^2 x_{23}^2}, \qquad\qquad v = \frac{x_{12}^2 x_{34}^2}{x_{13}^2 x_{24}^2}.
\end{equation}
This result straightforwardly generalizes to $n$-point functions, which now involve an arbitrary function of $n(n-3)/2$ cross ratios.

These results are easy to derive
in position space where the conformal group acts naturally.  
Yet for many modern applications,  including cosmology \cite{Antoniadis:2011ib, Maldacena:2011nz, Bzowski:2011ab, Kehagias:2012td, Kehagias:2012td, Bzowski:2012ih,Mata:2012bx, McFadden:2013ria, Ghosh:2014kba, Kundu:2014gxa, Anninos:2014lwa, Arkani-Hamed:2015bza, Isono:2016yyj, Arkani-Hamed:2018kmz, Arkani-Hamed:2017fdk, Anninos:2019nib, Sleight:2019mgd, Sleight:2019hfp}, 
condensed matter \cite{Chowdhury:2012km, Huh:2013vga,Jacobs:2015fiv, Lucas:2016fju, Myers:2016wsu, Lucas:2017dqa}, anomalies \cite{Coriano:2017mux, Gillioz:2018kwh, Coriano:2018zdo} and the bootstrap programme \cite{Polyakov:1974gs, Gopakumar:2016wkt, Gopakumar:2016cpb, Isono:2018rrb, Isono:2019wex}, 
it would be highly desirable to know the analogue of this result -- and, indeed, 
the analogue 
of the conformal cross-ratios themselves --  in \emph{momentum space}.

Despite the lapse of nearly five decades, such an understanding has yet to be achieved.
Nevertheless, through recent efforts, all the necessary prerequisites 
are now  in place.
Firstly,  the momentum-space 3-point functions of general scalar and tensorial operators are known, 
including the cases where 
anomalies and beta functions arise as a result of
renormalization
\cite{Armillis:2009pq, Coriano:2012wp, Bzowski:2013sza, Coriano:2013jba, Bzowski:2015pba, Bzowski:2017poo, Bzowski:2018fql, Coriano:2018bbe,  Gillioz:2018mto, Farrow:2018yni, Isono:2019ihz, Bautista:2019qxj,Gillioz:2019lgs}.  
Secondly, 
momentum-space studies of the 4-point function have yielded
special classes of solutions to the conformal Ward identities \cite{Raju:2012zs, Albayrak:2018tam, Isono:2018rrb, Li:2018wkt, Arkani-Hamed:2018kmz, Albayrak:2019asr, Maglio:2019grh}. 
Here, our aim is now to provide 
the {\it general} solution for the momentum-space 
$n$-point function.  We start by providing a complete discussion of the 4-point function and an exploration of its properties, and we then present the result for the $n$-point function.

\section{II.  \hspace{1mm} Momentum-space representation}

For scalar 4-point functions, 
our main result is the  
general momentum-space representation:
\begin{align}
& \lla \O_{\Delta_1}(\bs{p}_1) \O_{\Delta_2}(\bs{p}_2) \O_{\Delta_3}(\bs{p}_3) \O_{\Delta_4}(\bs{p}_4) \rra  \nn\\
& \quad = \int \frac{\D^d \bs{q}_{1}}{(2 \pi)^d} \frac{\D^d \bs{q}_{2}}{(2 \pi)^d} \frac{\D^d \bs{q}_{3}}{(2 \pi)^d} \frac{\hat{f}(\hat{u}, \hat{v})}{\text{Den}_3(\bs{q}_{j}, \bs{p}_k)}. \label{kint4}
\end{align}
Here,
$\<  \ldots \> = \lla  \ldots \rra (2\pi)^d\delta(\sum_i\bs{p}_i)
$,  $d$ is the spacetime dimension, and  
the denominator is 
\begin{align}
& \text{Den}_3(\bs{q}_{j}, \bs{p}_k) =q_{3}^{2 \delta_{12} + d} q_{2}^{2 \delta_{13} + d}  q_{1}^{2 \delta_{23} + d} 
|\bs{p}_1 + \bs{q}_{2} - \bs{q}_{3}|^{2 \delta_{14} + d}
 \nn\\& \quad \qquad \times |\bs{p}_2 + \bs{q}_{3} - \bs{q}_{1}|^{2 \delta_{24} + d} 
|\bs{p}_3 + \bs{q}_{1} - \bs{q}_{2}|^{2 \delta_{34} + d} \label{Den3}
\end{align}
where the $\delta_{ij}$ are given in  \eqref{params}.
We work in Euclidean signature throughout.
As expected from \eqref{x4point}, 
this 4-point function depends on 
an arbitrary 
function $\hat{f}(\hat{u},\hat{v})$ of two variables: 
\begin{align}
\hat{u} & = \frac{q_{1}^2 |\bs{p}_1 + \bs{q}_{2} - \bs{q}_{3}|^2}{q_{2}^2 |\bs{p}_2 + \bs{q}_{3} - \bs{q}_{1}|^2},\qquad  
\hat{v} = \frac{q_{2}^2 |\bs{p}_2 + \bs{q}_{3} - \bs{q}_{1}|^2}{q_{3}^2 |\bs{p}_3 + \bs{q}_{1} - \bs{q}_{2}|^2}.
\end{align}
The role of $\hat{u}$ and $\hat{v}$ is analogous to that of the position-space cross-ratios $u$ and $v$ defined in \eqref{std_uv}.  
These variables are thus the desired momentum-space cross-ratios, 
though notice they 
depend on the momenta $\bs{q}_j$ that are 
subject to integration in \eqref{kint4}.

\subsection{\hspace{-3mm}A.  \hspace{1mm}  Proof of conformal invariance}

The conformal invariance of \eqref{kint4} can be verified by direct substitution into the conformal Ward identities (CWIs). Its Poincar\'{e} invariance is manifest, and its scaling dimension is given by the sum of powers in \eqref{Den3} minus $3d$ from the three integrals. This gives $- 2 \delta_t - 3 d = \Delta_t - 3 d$, the correct result in momentum space.

The remaining CWIs associated with special conformal transformations are implemented by the second-order differential operator 
$\mathcal{K}^\kappa = \sum_{j=1}^3 \mathcal{K}^\kappa_j$, where  \cite{Bzowski:2013sza}
\begin{align}
\mathcal{K}^\kappa_j & =p_j^\kappa \frac{\partial}{\partial p_j^\alpha} \frac{\partial}{\partial p_{j}^{\alpha}} - 2 p_j^\alpha \frac{\partial}{\partial p_j^\alpha} \frac{\partial}{\partial p_j^\kappa} 
+ 2 (\Delta_j - d) \frac{\partial}{\partial p_j^\kappa}. \label{e:ward_cwi0}
\end{align}
By acting with 
$\mathcal{K}^\kappa$ on the integrand of \eqref{kint4}, one can show 
\begin{align}
& \mathcal{K}^\kappa \left(\frac{\hat{f}(\hat{u}, \hat{v})}{\text{Den}_3(\bs{q}_{j}, \bs{p}_k)} \right) =
\sum_{n=1}^3 \frac{\partial}{\partial q_n^\mu} \left[ \frac{(q_n)_{\alpha}}{\text{Den}_3(\bs{q}_{j}, \bs{p}_k)} \right. \nn\\
& \qquad \times \left. \left( \mathcal{A}_{(n)}^{\alpha\mu\kappa} \hat{u} \frac{\partial \hat{f}}{\partial \hat{u}} + \mathcal{B}_{(n)}^{\alpha\mu\kappa} \hat{v} \frac{\partial \hat{f}}{\partial \hat{v}} + \mathcal{C}_{(n)}^{\alpha\mu\kappa} \hat{f} \right) \right]. \label{Kon4}
\end{align}
In order 
to write these coefficients explicitly, we define
\begin{align}
A_{(n)}^{\alpha\mu\kappa} = \frac{2 k_n^\beta}{k_n^2} \left( \delta^{\kappa\alpha} \delta^\mu_\beta - \delta^{\mu\alpha} \delta^\kappa_\beta - \delta^{\mu\kappa} \delta^\alpha_\beta \right),
\end{align}
where the $\bs{k}_n$ are the vectors featuring in \eqref{Den3}, \textit{i.e.}, 
$\bs{k}_1 = \bs{p}_1 + \bs{q}_{2} - \bs{q}_{3}$ along with cyclic permutations. 
The coefficients in equation \eqref{Kon4} are then 
\begin{align}
\mathcal{C}_{(1)}^{\alpha\mu\kappa} & = ( \tfrac{d}{2} + \delta_{24}) A_{(2)}^{\alpha\mu\kappa} + ( \tfrac{d}{2} + \delta_{34}) A_{(3)}^{\alpha\mu\kappa},
\end{align}
with $C_{(2)}^{\alpha\mu\kappa}$ and $C_{(3)}^{\alpha\mu\kappa}$ following by cyclic permutation of the indices $1,2,3$, while
\begin{align}
& \mathcal{A}_{(1)}^{\alpha\mu\kappa} = A_{(2)}^{\alpha\mu\kappa}, && \mathcal{B}_{(1)}^{\alpha\mu\kappa} = A_{(3)}^{\alpha\mu\kappa} - A_{(2)}^{\alpha\mu\kappa}, \nn\\
& \mathcal{A}_{(2)}^{\alpha\mu\kappa} = - A_{(1)}^{\alpha\mu\kappa}, && \mathcal{B}_{(2)}^{\alpha\mu\kappa} = A_{(3)}^{\alpha\mu\kappa}, \nn\\
& \mathcal{A}_{(3)}^{\alpha\mu\kappa} = A_{(2)}^{\alpha\mu\kappa} - A_{(1)}^{\alpha\mu\kappa}, && \mathcal{B}_{(3)}^{\alpha\mu\kappa} = - A_{(2)}^{\alpha\mu\kappa}.
\end{align}
As the action of $\mathcal{K}^\kappa$  on the integrand of \eqref{kint4} is a total derivative, the integral itself is then invariant.  This proves the conformal invariance of the representation \eqref{kint4}.

\subsection{\hspace{-3mm}B.  \hspace{1mm}  The tetrahedron} 

\begin{figure}[ht]
\hspace{-1.1cm}\includegraphics[width=6.0cm]{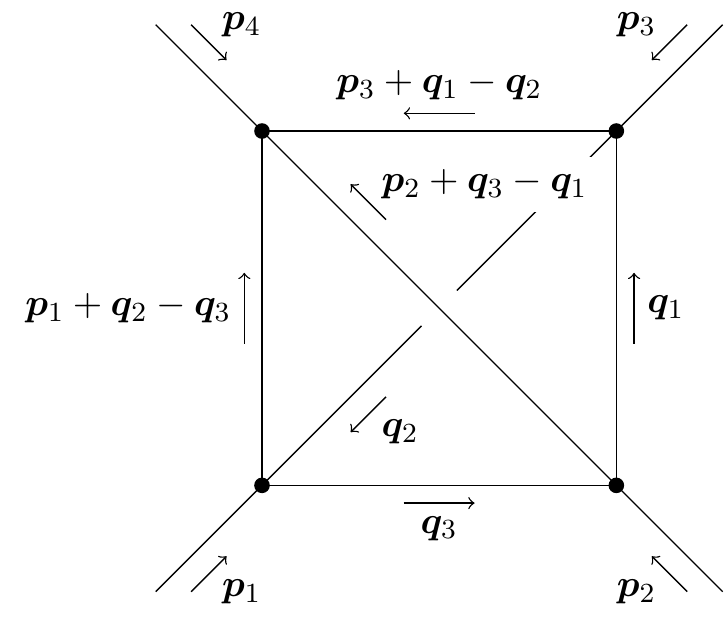} 
\centering
\caption{The 3-loop tetrahedral integral \eqref{intmono}, where  each internal line corresponds to a generalized propagator in \eqref{Den3ab}. 
 \label{fig:1}}
\end{figure}

The momentum-space expression \eqref{kint4} is not the direct Fourier transform of the position-space expression \eqref{x4point}.  Rather, for $f(u,v) = u^{\alpha} v^{\beta}$, the Fourier transform is given by \eqref{kint4} with
\begin{align} \label{monof}
\hat{f}(\hat{u}, \hat{v}) = C^{\delta_{12}, \delta_{34}}_{\beta} C^{\delta_{13}, \delta_{24}}_{\alpha - \beta} C_{- \alpha}^{\delta_{14}, \delta_{23}} \hat{u}^\alpha \hat{v}^\beta,
\end{align}
where
\begin{align}
C^{\delta, \delta'}_{\sigma} = 4^{\delta+\delta'+2 \sigma + d} \pi^{d} \frac{\Gamma \left( \frac{d}{2} + \delta + \sigma \right) \Gamma \left( \frac{d}{2} + \delta' + \sigma \right)}{\Gamma(-\delta-\sigma) \Gamma(-\delta'-\sigma)}.
\label{Cab}
\end{align}
This follows since 
the Fourier transform of a product is a convolution of Fourier transforms, and so we can write
\begin{align}
& \mathcal{F} \left[ x_{12}^{2(\beta + \delta_{12})} x_{34}^{2(\beta + \delta_{34})} \times  x_{13}^{2(\alpha-\beta + \delta_{13})} x_{24}^{2(\alpha-\beta + \delta_{24})} \right.\nn\\ 
& \qquad\qquad \left. \times \: x_{14}^{2(- \alpha + \delta_{14})} x_{23}^{2(-\alpha + \delta_{23})} \right] \nn\\
&  = \mathcal{F}[x_{12}^{2(\beta + \delta_{12})} x_{34}^{2(\beta + \delta_{34})}] \ast \mathcal{F}[x_{13}^{2(\alpha-\beta + \delta_{13})} x_{24}^{2(\alpha-\beta + \delta_{24})}] \nn\\
& \qquad\qquad \ast \mathcal{F}[x_{14}^{2(- \alpha + \delta_{14})} x_{23}^{2(-\alpha + \delta_{23})}], \label{fullFT}
\end{align}
where $\ast$ denotes the convolution in all variables, namely $(f \ast g)(\bs{p}_k) = \int \prod_{j=1}^4 \frac{\D^d \bs{q}_j}{(2 \pi)^d} f(\bs{q}_j) g(\bs{p}_j - \bs{q}_j)$. With the $\hat{f}$  in \eqref{monof}, the momentum-space integral in \eqref{kint4} becomes 
\begin{align} \label{intmono}
W_{\alpha,\beta}=\int \frac{\D^d \bs{q}_{1}}{(2 \pi)^d} \frac{\D^d \bs{q}_{2}}{(2 \pi)^d} \frac{\D^d \bs{q}_{3}}{(2 \pi)^d} \frac{1}{\text{Den}^{(\alpha \beta)}_3(\bs{q}_{j}, \bs{p}_k)},
\end{align}
where 
\begin{align} \label{Den3ab}
& \text{Den}_3^{(\alpha \beta)}(\bs{q}_{j}, \bs{p}_k) = q_{3}^{2\gamma_{12}} q_{2}^{2 \gamma_{13}}  q_{1}^{2 \gamma_{23}}  |\bs{p}_1 + \bs{q}_{2} - \bs{q}_{3}|^{2 \gamma_{14}} 
\nn\\
& \qquad \times
 |\bs{p}_2 + \bs{q}_{3} - \bs{q}_{1}|^{2 \gamma_{24}}  |\bs{p}_3 + \bs{q}_{1} - \bs{q}_{2}|^{2 \gamma_{34}}
\end{align}
and
\begin{align}\label{gammadef}
\gamma_{12} &=  \delta_{12} + \beta + d/2,
&&\gamma_{13} =\delta_{13} + \alpha - \beta + d/2,\nn \\
\gamma_{23}  &= \delta_{23} - \alpha + d/2, 
&& \gamma_{14} =\delta_{14} - \alpha + d/2, \nn\\
\gamma_{24} &=\delta_{24} + \alpha - \beta + d/2,
&& \gamma_{34}= \delta_{34} + \beta + d/2.
\end{align}
This is a 3-loop Feynman integral with the topology of a tetrahedron as presented in Fig.~\ref{fig:1}. 
The four momenta entering the vertices are those of the external operators, while the six internal lines describe generalized propagators in which the momenta are raised to the specific powers given in \eqref{gammadef}.

\subsection{\hspace{-3mm}C.  \hspace{1mm} Spectral representation}

Where convergence permits, the function $\hat{f}(\hat{u},\hat{v})$ can be expressed as a double inverse Mellin transform over the monomial  $\hat{u}^\alpha\hat{v}^{\beta}$. 
The 4-point function \eqref{kint4}  then admits the  spectral   representation
\begin{align}\label{Mellint}
& \lla \O_{\Delta_1}(\bs{p}_1) \O_{\Delta_2}(\bs{p}_2) \O_{\Delta_3}(\bs{p}_3) \O_{\Delta_4}(\bs{p}_4) \rra \nn\\&\quad = 
\frac{1}{(2\pi i)^2}
\int_{b_1-i\infty}^{b_1+i\infty}\D\alpha\int_{b_2-i\infty}^{b_2+i\infty}\D\beta \,\rho(\alpha,\beta)\,W_{\alpha,\beta}
\end{align}
for an appropriate choice of integration contour specified by $b_1$ and $b_2$.
Here, $W_{\alpha,\beta}$ is a universal kernel corresponding to the tetrahedron integral \eqref{intmono} and 
$\rho(\alpha,\beta)$ is a theory-specific spectral function derived from the Mellin transform of $\hat{f}(\hat{u},\hat{v})$.
Where the {\it position-space} Mellin representation of a 4-point function is known -- as is often the case for holographic CFTs \cite{Mack:2009mi, Penedones:2010ue, Fitzpatrick:2011ia} --  the corresponding $\rho(\alpha,\beta)$ in momentum space can  be read off immediately using  equations 
\eqref{monof} and \eqref{Cab}. 

\label{sec:simpofres}

\begin{figure}[ht]
\hspace{-0.45cm}
\includegraphics[width=0.50\textwidth]{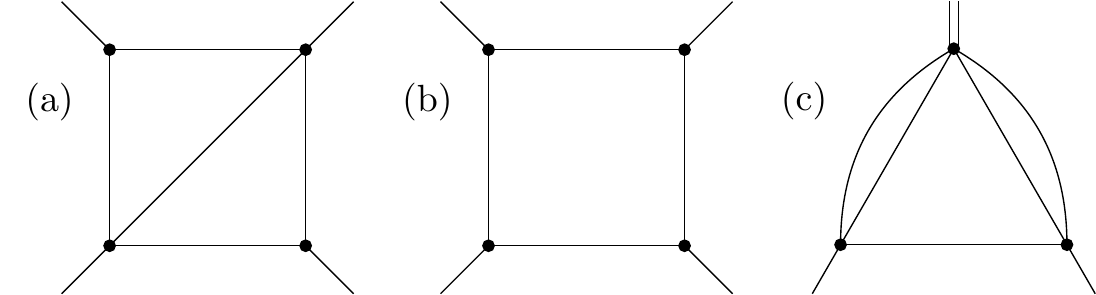} 
\centering
\caption{
Simplifications of the kernel $W_{\alpha,\beta}$: 
(a) where a propagator in \eqref{Den3ab} appears with $\gamma_{ij}=d/2+n$ the loop order is reduced by one;
(b) with two such propagators 
we obtain a 1-loop box;
(c) for  $\gamma_{ij}=-n$, we obtain a 3-point function.
 \label{simpfig}}
\end{figure}

To evaluate the spectral integral, we close the contour and sum the residues.
For certain 
$\alpha$ and $\beta$, these residues are 
 simple to evaluate due to reductions in the loop order of $W_{\alpha,\beta}$.
Such reductions 
arise whenever a propagator in the denominator \eqref{Den3ab} appears with a power $\gamma_{ij} = d/2+n$, for some non-negative integer $n$.   This can be seen by noting that, in a distributional sense as $q\rightarrow 0$,
\begin{equation} \label{delta}
 \lim_{\epsilon \rightarrow 0} \, \frac{\epsilon}{q^{d+2n-2 \epsilon}} = \frac{\pi^{d/2}}{4^n n! \Gamma(d/2+n)}\Box^n\delta^{(d)}(\bs{q}).
 \end{equation}
We then obtain a pole in $\alpha$, $\beta$ whose residue is given by a 2-loop integral 
as shown in Fig.~\ref{simpfig}{\color{navy}a}.  
Where the external dimensions permit, such poles can also coincide.   In Fig.~\ref{simpfig}{\color{navy}b}, we illustrate the case where $\alpha-\beta=\delta_{14} = \delta_{23}$, creating a  pair of delta functions $\delta(\bs{q}_2) \delta(\bs{p}_2 + \bs{q}_3 - \bs{q}_1)$ for which the  residue is  a 1-loop box.

Simplifications of a different kind occur whenever a propagator in \eqref{Den3ab} appears with a vanishing power, or more generally for $\gamma_{ij}=-n$.  This results in a contraction of the corresponding leg of the tetrahedron, producing a 1-loop triangle for which two of the legs are bubbles as shown in Fig.~\ref{simpfig}c.  Evaluating the bubbles, one obtains a pure 1-loop triangle 
whose propagators are raised to new powers.  This integral is equivalent to a general CFT 3-point function \cite{Bzowski:2013sza}.
The locality can be understood by noting that 
a denominator $q^{-2n}$ corresponds to a factor $x_{ij}^{-(d+2n)}$ in position space, which is equivalent to a delta function via the position-space analogue of \eqref{delta}.

\begin{figure*}[ht]
\includegraphics[width=0.75\textwidth]{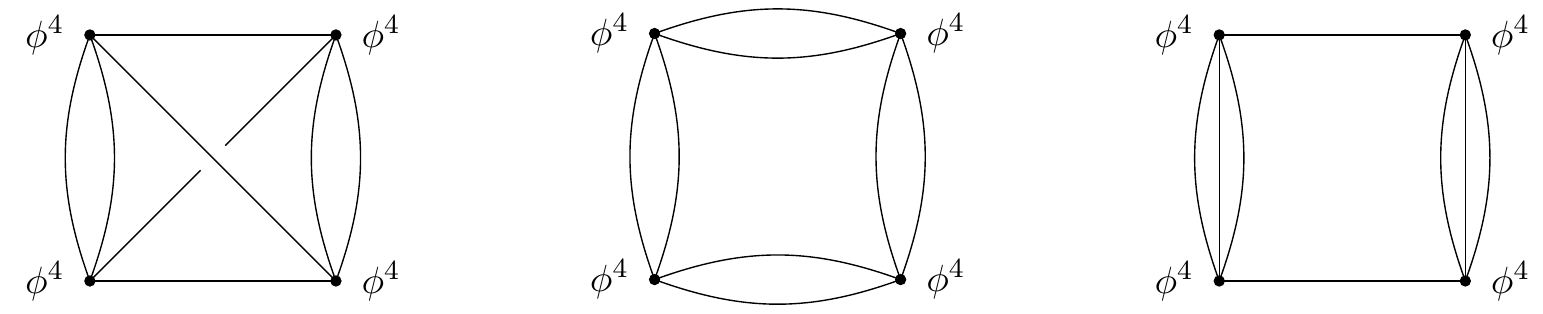}
\centering
\caption{Three 
distinct topologies of Feynman diagrams contributing to the connected part of  $\< \lwick \phi^4 \rwick \lwick \phi^4 \rwick \lwick \phi^4 \rwick \lwick \phi^4 \rwick \>$.\label{fig:2}}
\end{figure*}

\subsection{\hspace{-3mm}D.  \hspace{1mm} Singularities and renormalization}
\label{Sec:singularities}

For special values of the spacetime and operator dimensions, momentum-space CFT correlators exhibit divergences requiring regularization and renormalization.   
All divergences are local can be removed 
 through the addition of  covariant counterterms giving rise to conformal anomalies and beta functions for composite operators.  
The renomalization of 3-point functions was studied in \cite{Bzowski:2015pba, Bzowski:2017poo, Bzowski:2018fql}.  For 4-point functions, a similar analysis holds as we now discuss. 

Firstly, renormalizability requires that all UV divergences should be either {\it ultralocal}, with support only when all four position-space insertions are coincident, or else {\it semilocal}, meaning they are supported only in the cases where either (i) $\bs{x}_1 = \bs{x}_2 = \bs{x}_3 \neq \bs{x}_4$, (ii) $\bs{x}_1 = \bs{x}_2 \neq \bs{x}_3 = \bs{x}_4$, or (iii) $\bs{x}_1 = \bs{x}_2 \neq \bs{x}_3 \neq \bs{x}_4$, along with all related cases obtained by permutation.  
In momentum space, ultralocal divergences are thus analytic in all the squared momenta, while semilocal divergences are analytic in at least one  squared momentum.   
(Cases (i) and (ii) have a momentum-dependence matching that of a 2-point function, while that of case (iii) corresponds to a 3-point function.)

These divergences constitute local solutions of the CWI. 
  Their form,  
as well as the $d$ and $\Delta_j$ 
for which they appear, 
can be predicted from an analysis of local counterterms. 
Such counterterms 
exist only in cases where
\begin{align}\label{singcond}
d + \sum_{j=1}^4 \sigma_j (\Delta_j-d/2) = -2 n
\end{align}
for some $n$ non-negative integer,  with signs $\sigma_j$  whose values are either all minus, or else three minus and one plus.

{\em Ultralocal divergences}  
are removed by counterterms that are quartic in the sources $\varphi_j$ for the operators $\O_{\Delta_j}$.  
These feature  $2n$ fully-contracted derivatives whose action is distributed over the sources, and exist whenever 
\eqref{singcond} is satisfied with all minus signs. 
Since the scaling  dimension of $\varphi_j$ is $d-\Delta_j$, this 
ensures 
the counterterm has overall dimension $d$.
The appearance of ultralocal divergences when 
this 
condition 
is satisfied 
can be seen by examining the region of integration where all three loop momenta in the kernel $W_{\alpha,\beta}$ become large simultaneously.  
Re-parametrizing $\bs{q}_j = \lambda \hat{\bs{q}}_j$ where $\hat{q}_1^2=1$, as $\lambda\rightarrow\infty$ the denominator in \eqref{intmono} scales as $\lambda^{6d-\Delta_t}(1+O(\lambda^{-2}))$, while the numerator contributes a Jacobian factor  $\int \D\lambda\,\lambda^{3d-1}$.  The $\lambda$ integral is then logarithmically divergent precisely when \eqref{singcond} is satisfied with all minus signs.  (For nonzero $n$, the divergence derives from expanding the denominator to subleading order in powers of $\lambda^{-2}$.)
After the divergence is subtracted and the regulator removed, the renormalized correlator has the expected nonlocal momentum dependence and obeys anomalous CWIs due to the RG scale introduced by the counterterm, see \cite{Bzowski:2015pba}.

{\em Semilocal divergences} 
are removed by counterterms featuring one operator and multiple
sources.  
For quartic counterterms, we have  $2n$ fully-contracted derivatives whose action is distributed over $\varphi_1\varphi_2
 \varphi_3\O_{\Delta_4}$.  Such counterterms exist whenever \eqref{singcond} is satisfied with signs $(---\,+)$ (or some permutation thereof), 
ensuring the counterterm has dimension $d$.  The resulting 4-point contribution then has the momentum-dependence of a 2-point function and  corresponds to case (i) above.
This counterterm effectively reparametrizes the source for $\O_{\Delta_4}$ and we obtain a  beta function 
 in the renormalized theory. 

The appearance of a semilocal divergence in $W_{\alpha,\beta}$ when the $(---\,+)$ condition 
is satisfied can be seen by re-parametrizing the loop momenta in \eqref{intmono} as $\bs{q}_1 = \lambda \hat{\bs{q}}_1$ with $\hat{q}_1^2=1$, $\bs{q}_2 = \lambda \hat{\bs{q}}_1+\bs{p}_3+\bs{\ell}_2$ and $\bs{q}_3 = \lambda \hat{\bs{q}}_1-\bs{p}_2+\bs{\ell}_3$.   
The $\lambda$ integral is then logarithmically divergent when 
this condition is satisfied, and has a semilocal momentum dependence that is non-analytic in $p_4^2$ only.
For the permuted cases featuring $\O_{\Delta_j}$ with $j=1,2,3$ in place of $\O_{\Delta_4}$, the corresponding reparametrization is simply $\bs{q}_j=\lambda\hat{\bs{q}}_j$ leaving the other loop momenta fixed.  This difference  
reflects our use of momentum conservation to eliminate $\bs{p}_4$ in 
\eqref{intmono}.
After renormalization, the correlator is again fully nonlocal and obeys anomalous CWIs reflecting the presence of the beta function, see \cite{Bzowski:2015pba}.

Besides the quartic counterterms discussed above, which contribute solely to 4- and higher-point functions, 
we may also have 
cubic and quadratic counterterms.  
Their form 
is already fixed from the renormalization of  2- and 3-point functions  \cite{Petkou:1999fv, Bzowski:2015pba, Bzowski:2017poo, Bzowski:2018fql}, but they nevertheless  
contribute to 4-point functions as well \cite{Bzowski:2016kni}.
In particular, cubic counterterms with two sources and one operator 
remove semilocal divergences of types (ii) and (iii).

\section{\hspace{-3mm}III.  \hspace{1mm} Free fields}

Consider a free spin-0 massless field $\phi$ and connected 4-point functions of the operators of the form $\phi^n$. In all cases, the function $f$ in position space is a sum of monomials of the form $u^\alpha v^\beta$. For example, for the connected 4-point function $ \< \lwick \phi^2 \rwick \lwick \phi^2 \rwick \lwick \phi^2 \rwick \lwick \phi^2 \rwick \>_{\text{conn}} $, one has
\begin{equation}
f(u,v) \sim \left( \frac{u}{v} \right)^{\frac{1}{6} \Delta_{\phi^2}} + \left( \frac{v}{w} \right)^{\frac{1}{6} \Delta_{\phi^2}} + \left( \frac{w}{u} \right)^{\frac{1}{6} \Delta_{\phi^2}}
\end{equation}
where $\Delta_{\phi^2} = d - 2$ is the dimension of $\phi^2$, and
to write $f$ and $\hat{f}$ succinctly we introduce the additional conformal ratios $w$ and $\hat{w}$ defined by  
 $u v w = 1$ and $\hat{u} \hat{v} \hat{w} = 1$. 
Equation \eqref{monof} now yields the momentum space $\hat{f}$. In this case, however, the prefactor in \eqref{monof} vanishes as two out of the six gamma functions in the denominator of \eqref{Cab} diverge. This means we have to consider the regulated expression with regulated $\hat{f}$, namely
\begin{equation} \label{phi2}
\hat{f}(\hat{u},\hat{v}) = 16 \tilde{\epsilon}^2 \left( \frac{\hat{u}}{\hat{v}} \right)^{\frac{1}{6} \Delta_{\phi^2} - \frac{1}{2} \epsilon} + 2 \text{ cycl.\,perms.},
\end{equation}
where $\tilde{\epsilon}=\epsilon\,(4\pi)^{d/2}\Gamma(d/2)$ and $2 \text{ cycl.\,perms.}$ denotes two remaining terms with cyclic permutations of the ratios, $\hat{u} \mapsto \hat{v} \mapsto \hat{w} \mapsto \hat{u}$. After this is substituted into \eqref{kint4} and the momentum space integrals carried out, the limit $\epsilon \rightarrow 0$ should be taken.

The appearance of the double zero in \eqref{phi2} reflects the fact that the only Feynman diagram contributing to this correlator has the topology of a box. If instead we consider the 4-point function of $\lwick \phi^4 \rwick$, the contributing Feynman diagram topologies are as  presented in Fig.~\ref{fig:2}. Up to an overall symmetry factor, the regulated $\hat{f}$ reads
\begin{align}
\hat{f}(\hat{u},\hat{v}) & \sim \left( c_2^2 \left( \frac{\hat{v}}{\hat{u}} \right)^{\frac{1}{12} \Delta_{\phi^4}} + \tilde{\epsilon}^2 c_2^4 \left( \frac{\hat{u}}{\hat{v}} \right)^{\frac{1}{6} \Delta_{\phi^4} - \frac{1}{2} \epsilon} + \right.\nn\\
& \left. + \: \tilde{\epsilon}^2 c_3^2 \left(\frac{\hat{v}^4}{\hat{u}} \right)^{\frac{1}{12} \Delta_{\phi^4} - \frac{1}{4} \epsilon} \right) + 2 \text{ cycl.\,perms.},
\end{align}
where $\Delta_{\phi^4} = 2(d-2)$. The constants $c_n$ are defined recursively through
\begin{equation} \label{cns}
c_{n+1} = c_n \, \frac{\Gamma(\Delta_{\phi}) \Gamma(n \Delta_{\phi}) \Gamma(1 - n \Delta_{\phi})}{(4 \pi)^{d/2} \Gamma( (n+1) \Delta_{\phi}) \Gamma(1 - (n-1) \Delta_{\phi})}
\end{equation}
with $c_1 = 1$ and $\Delta_{\phi} = d/2 - 1$. These coefficients arise from the evaluation of effective propagators. Denoting the standard massless propagator as $D_1(p) = 1/p^2$, the effective propagator $D_n(p)$ with $n$ lines in Fig.~\ref{fig:2} is
\begin{align}
D_n(p) & = \frac{c_n}{p^{2 - 2 (n-1) \Delta_{\phi}}} = \int \frac{\D^{d} \bs{q}}{(2 \pi)^{d}} \frac{D_{n-1}(q)}{|\bs{p} - \bs{q}|^2}.
\end{align}

Finally, the disconnected part of any correlator can also be represented by the function $\hat{f}$. As an example, consider a generalized free field $\O$ of dimension $\Delta_{\O}$, for which the position-space 4-point function has
\begin{equation}
f(u,v) \sim \left( \frac{v}{u} \right)^{\frac{1}{3} \Delta_{\O}} + 2 \text{ cycl.\,perms.}
\end{equation}
The momentum-space expression is then proportional to $p_1^{2 \Delta_{\O} - d} p_3^{2 \Delta_{\O} - d} \delta(\bs{p}_1 + \bs{p}_2) \delta(\bs{p}_3 + \bs{p}_4)$ plus permutations, and can be represented by a regulated $\hat{f}$ with quadruple zero,
\begin{equation}
\hat{f}(\hat{u},\hat{v}) = \tilde{\epsilon}^4 \left[ \left( \frac{\hat{v}}{\hat{u}} \right)^{\frac{1}{3} \Delta_{\O} - \epsilon} + 2 \text{ cycl.\,perms.} \right].
\end{equation}

\section{\hspace{-3mm}IV.  \hspace{1mm} Holographic CFTs}

Holographic 4-point functions are obtained by evaluating Witten diagrams in anti-de Sitter space.  
These yield compact scalar integral representations for the momentum-space 4-point functions.
Such expressions must again be special cases of the general solution \eqref{kint4} for some appropriate $\hat{f}$. 
This function can be found in several ways as we now discuss.
Since exchange Witten diagrams can be reduced to a sum of contact diagrams \cite{DHoker:1999mqo, Dolan:2000ut}, 
we focus here on the latter deferring a complete discussion to \cite{Long}.
In the simplest case of a quartic bulk interaction without derivatives, we find 
\begin{align}
\label{4Kint}
 \Phi &=\lla \O_{\Delta_1}(\bs{p}_1) \O_{\Delta_2}(\bs{p}_2) \O_{\Delta_3}
(\bs{p}_3) \O_{\Delta_4}(\bs{p}_4) \rra   \nn\\&
= c_W\int_0^\infty \D z\, z^{d-1}\prod_{j=1}^4 p_j^{\Delta_j-d/2}K_{\Delta_j-d/2}(p_j z),
\end{align}
where $c_W= 2^{2d+4-\Delta_t}/\prod_{j=1}^4\Gamma(\Delta_j-d/2)$ and the four modified Bessel-$K$ functions represent bulk-boundary propagators.  

\begin{figure}
\hspace{-0.25cm}\includegraphics[width=0.4\textwidth]{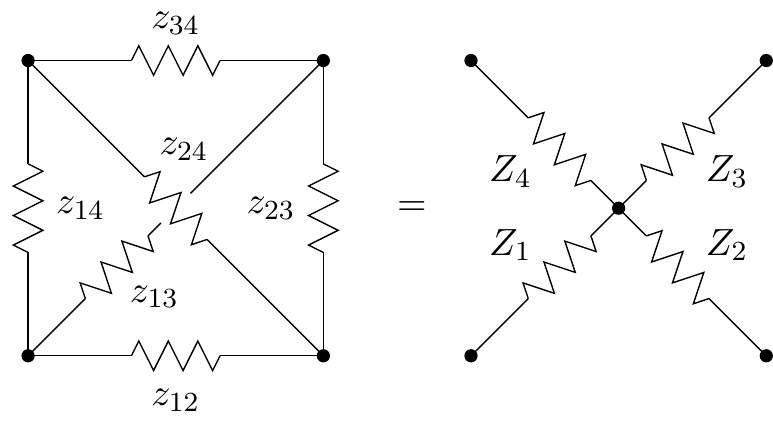}
\caption{Equivalent electrical circuits where the impedances are related by $z_{ij}=Z_iZ_j/Z_t$.  
Setting $z_{i4}=s_i$ for $i=1,2,3$ and $(z_{12},z_{23},z_{31})=(z^2/s_3,z^2/s_1,z^2/s_2)$ gives a mapping of Schwinger parameters converting the contact Witten diagram \eqref{impedancerep} into the form \eqref{kint4} with $\hat{f}(\hat{u},\hat{v})$ given in \eqref{fhatcontact}.
 \label{fig:4}}
\end{figure}

This integral can now be mapped to a tetrahedral topology via  the star-mesh transformation from electrical circuit theory.   
Schwinger parametrizing the Bessel functions in \eqref{4Kint} and evaluating the integral,  
we find
\begin{align}\label{impedancerep}
\Phi=c'_W\prod_{j=1}^4 \int_0^\infty\D Z_j \,Z_j^{\Delta_j-d/2-1} Z_t^{(d-\Delta_t)/2} e^{-p_j^2/2Z_j}
\end{align}
for $c_W' = 2^{(\Delta_t-d)/2-5}\Gamma(\frac{1}{2}(\Delta_t{-}d)) c_W$ and  $Z_t = \sum_{j=1}^4 Z_j$.  The exponent describes the power dissipated in a network of four impedances $Z_j$ 
arranged in a star configuration.  Such a  network is equivalent, however, to a  tetrahedral network where the impedance connecting the vertices $(i,j)$ is $z_{ij}=Z_iZ_j/Z_t$ (see Fig.~\ref{fig:4}).  Since all products of the impedances on  opposite edges are equal,  $ z^2 = z_{12}z_{34}=z_{13}z_{24}=z_{14}z_{23}$, we can re-parametrise the  tetrahedron in terms of $z$ and the three variables $s_i=z_{i4}$ for $i=1,2,3$.  
With this change of variables, the contact diagram \eqref{4Kint} can be mapped to the form \eqref{kint4}, with
\begin{align}\label{fhatcontact}
&\hat{f}(\hat{u},\hat{v})=16 c'_W (2\pi)^{3d/2}\Big(\frac{\hat{u}}{\hat{v}}\Big)^{-\Delta_t/12+d/2}
\nn\\ &\qquad \times
\int_0^\infty \D z 
\,z^{-\Delta_t/2+3d-1}   
K_{\delta_{13}-\delta_{24}}(z)
\nn\\ & \qquad\qquad  \times
 K_{\delta_{23}-\delta_{14}}(z\sqrt{\hat{u}})K_{\delta_{12}-\delta_{34}}(z/\sqrt{\hat{v}}).
\end{align}
This can be directly verified by 
Schwinger parametrizing the three Bessel functions in terms of the $s_i$ then performing the Gaussian integrations over the  momenta $\bs{q}_i$ in \eqref{kint4}. (For full details, see \cite{Long}).  
Remarkably, this $\hat{f}$ features precisely the same integral (the 
`triple-$K$')   
that describes
 the  momentum-space {\it 3-point} function \cite{Bzowski:2013sza}.  

An alternative derivation of \eqref{fhatcontact} starts from the position-space Mellin representation for the contact Witten diagram \cite{Penedones:2010ue}.  Applying \eqref{Cab}, one immediately obtains a spectral representation of the form \eqref{Mellint} with
\begin{align}\label{contactMellin}
&\rho(\alpha,\beta) =
c'_W 2^{-\Delta_t/2+3d}(2\pi)^{3d/2}  \prod_{i<j}\Gamma(\gamma_{ij})
\end{align}
and the $\gamma_{ij}$ defined in \eqref{gammadef}.
The equivalence of this result with \eqref{fhatcontact} is seen by writing the latter as a double inverse Mellin transform.
The poles of this spectral function now give residues of $W_{\alpha,\beta}$ for which the propagators in \eqref{Den3ab} have powers $\gamma_{ij}=-n$.  The ensuing reduction to 3-point functions shown in  
Fig.~\ref{simpfig}{\color{navy}c} then accounts for the appearance of the triple-$K$ integral in \eqref{fhatcontact}.
It would be interesting to understand if this simplification of residues is a general feature of holographic 4-point functions.

\section{\hspace{-3mm}V. \hspace{1mm} n-point function}

Generalizing our discussion above, the conformal $n$-point function takes the form 
\cite{Long}
\begin{align} \label{int1a}
&\< \O_1(\bs{p}_1) \ldots \O_n(\bs{p}_n) \> \\&=
 \prod_{1 \leq i < j \leq n} \int \frac{\D^d \bs{q}_{ij}}{(2 \pi)^d} \frac{\hat{f}(\{\hat{u}\})}{q_{ij}^{2 \delta_{ij} + d}}  \prod_{k=1}^n (2\pi)^d \delta \left( \bs{p}_k - \sum_{l=1}^n \bs{q}_{kl} \right),\nn
\end{align}
where  $\sum_{\vphantom{[} 1 \leq i < j \leq n} 2 \delta_{ij} = - \Delta_t$ and
$\hat{f}$ is an arbitrary function of $n(n-3)/2$ `conformal ratios' which we denote collectively as $\{ \hat{u}\}\vphantom{\big[}=q^2_{ij} q^2_{kl}/q^2_{ik} q^2_{jl}.
$
The tetrahedron thus generalizes to an $\vphantom{\big[}(n-1)$-simplex where $\bs{q}_{ij}$ is the momentum running from vertex $i$ to $j$.  We then have $n(n-1)/2$ integrals and $n-1$ delta functions (setting one aside for overall momentum conservation), leaving $(n-1)(n-2)/2$ integrals to perform.
If $n=4$, integrating out the delta functions and using $\bs{q}_a=\epsilon_{abc}\, \bs{q}_{bc}$, where $a, b, c=1, 2, 3$ and $\epsilon_{abc}$ is the Levi-Civita symbol,  we recover \eqref{kint4}.

\section{\hspace{-3mm}VI.  \hspace{1mm}  Conclusions}

We have presented a general momentum-space representation for the scalar $n$-point function of any CFT.  
This  features an arbitrary function of $n(n-3)/2$ variables  which play the role of momentum-space conformal ratios,
and is a solution of 
the conformal Ward identities.  It would be interesting to generalize this to tensorial correlators.

Following the success of the conformal bootstrap program in position space \cite{Simmons-Duffin:2016gjk, Poland:2018epd}, it 
may prove useful 
 to develop a  version in momentum space,  see \cite{Polyakov:1974gs, Gopakumar:2016wkt, Gopakumar:2016cpb, Isono:2018rrb, Isono:2019wex}.
 This requires understanding the expansion of the 4-point function in conformal partial waves
\cite{Ferrara:1973vz, Ferrara:1973yt, Polyakov:1974gs, Dolan:2003hv, Dolan:2011dv}.   
One then seeks to impose consistency with the  operator product expansion (OPE).  
To correctly implement the OPE in momentum space
requires a careful treatment of the short-distance singularities \cite{Bzowski:2014qja}.
To understand these better, and for practical calculational purposes, it would be 
useful to find a compact scalar parametric representation of the general 
solutions \eqref{kint4} and \eqref{int1a}.  For 3-point functions this is provided by the triple-$K$ integral, while for holographic $n$-point functions we have Witten diagrams. 
This suggests the existence of a similarly compact scalar representation for the general CFT $n$-point function.  
We hope to report on these questions in the near future.

\vspace{3mm}

\begin{acknowledgments}
{\em Acknowledgments.} 
AB is supported by the Knut and Alice Wallenberg Foundation under Grant No.~113410212.
PLM is supported by the Science and Technology Facilities Council through an Ernest Rutherford Fellowship No.~ST/P004326/2.  
 KS is supported in part by the Science and Technology Facilities Council (Consolidated Grant “Exploring the Limits of the Standard Model and Beyond”).
 
\end{acknowledgments} 
 

%

\end{document}